# Water Vapor Absorption in the Clear Atmosphere of an exo-Neptune

Jonathan Fraine[1,2,3], Drake Deming[1,4], Bjorn Benneke[3], Heather Knutson[3], Andrés Jordán[2], Néstor Espinoza[2], Nikku Madhusudhan[5], Ashlee Wilkins[1], Kamen Todorov[6]

(1) University of Maryland, Department of Astronomy
(2) Pontificia Universidad Católica de Chile, Instituto de Astrofísica
(3) California Institute of Technology, Division of Geological & Planetary Sciences
(4) NASA Astrobiology Institute's Virtual Planetary Laboratory
(5) University of Cambridge, Institute of Astronomy
(6) ETH Zürich, Department of Physics

**Transmission spectroscopy to date has detected atomic and molecular absorption in Jupiter-sized exoplanets, but intense efforts to measure molecular absorption in the atmospheres of smaller (Neptune-sized) planets during transits have revealed only featureless spectra[1,2,3,4]. From this it was concluded that the majority of small, warm planets evolve to sustain high mean molecular weights, opaque clouds, or scattering hazes in their atmospheres, obscuring our ability to observe the composition of these atmospheres[1,2,3,4,5]. Here we report observations of the transmission spectrum of HAT-P-11b (~4 Earth radii) from the optical to the infrared. We detected water vapour absorption at 1.4 micrometre wavelength. The amplitude of the water absorption (approximately 250 parts-per- million) indicates that the planetary atmosphere is predominantly clear down to ~1 mbar, and sufficiently hydrogen-rich to exhibit a large scale height. The spectrum is indicative of a planetary atmosphere with an upper limit of ~700 times the abundance of heavy elements relative to solar. This is in good agreement with the core accretion theory of planet formation, in which gas giant planets acquire their atmospheres by directly accreting hydrogen-rich gas from the protoplanetary nebulae onto a large rocky or icy core[6].**

We observed transits of HAT-P-11b[7] ($M_p = 25.8 \pm 2.9\ M_\oplus$; $R_p = 4.37 \pm 0.08\ R_\oplus$; $T_{eq} = 878 \pm 50$ K) in a joint Hubble-Spitzer program. Our Hubble observations comprised 1.1-1.7 micrometre grism spectroscopy using the Wide Field Camera 3 (WFC3) in spatial scanning mode. We also integrated these data over wavelength to produce WFC3 photometry[1,2,3,4,8]. Our Spitzer observations comprised photometry during two transits per 3.6 & 4.5 micrometre band of the IRAC instrument[9]. Because the planet lies in the Kepler field[10], precision optical photometry (~642nm) was obtained simultaneously with our Spitzer observations, although not simultaneously with our Hubble observations. Table 1 summarizes specific details of our observations and Figure 1a shows our transit photometry and model fits. Because HAT-P-11 is an active planet-hosting star[11,12,13], we show that starspots on the stellar surface are not sufficiently cool, nor sufficiently prevalent, to mimic the effect of water vapour absorption in the planet[14]. Our simultaneous Spitzer and Kepler photometry was critical to defining the temperature of the starspots that could otherwise, potentially mimic the effect of water vapour absorption in the planetary atmosphere.

HAT-P-11b crosses starspots on virtually every transit [12,13], as seen prominently Figure 1a. Our WFC3 photometry has the sensitivity to detect starspot crossings[2], but none were observed when HST observed the system. Our WFC3 observations contain large temporal gaps because Hubble passes behind the Earth[1,2,3,4,8,15], but not during the transit. Therefore, *un*occulted starspots, rather than occulted ones, potentially affect our transmission spectrum[16,17]. When the planet blocks unspotted portions of the stellar photosphere, the absorption lines in cool unoccluted spots become relatively more prominent[12,13].

Figure 1a shows the binned and normalised light curves of our four simultaneous Kepler-Spitzer transits and our WFC3 band-integrated light curve. We fit analytic transit light curves to all of time series with PyMC[18] to generate Markov Chain Monte Carlo (MCMC) distributions to estimate of the planetary parameters[19,20]. We re-analyzed the phased & binned Kepler data using improved limb darkening coefficients derived from stellar model atmospheres[21]. To fit the Spitzer and WFC3 transits, we hold the orbital distance and inclination constant at our Kepler-derived values. Although the uncertainties for the Kepler derived parameters were improved compared to previous studies[12,13], our purpose was to implement the updated limb darkening law and derive orbital parameters for all of our observations.

Each of the Kepler light curves obtained concurrently with our Spitzer observations show starspot crossings as deviations in the light curves between ~0.3-0.7 hours after mid-transit. The amplitude of these deviations is a function of both the area and temperature of the occulted spots [12,13]. Because the Kepler and Spitzer photometry were concurrent, the relative intensity is independent of the starspots' areas. On the other hand, because the starspots' temperature contrasts with the photosphere is a chromatic effect, the amplitude of these deviations varied with wavelength[16,17]. The spot crossings are not obvious in the Spitzer data because thermal radiation produces a much smaller contrast between the stellar photosphere and spot fluxes in the infrared than in the optical. The ratio between the Spitzer and Kepler spot crossing amplitudes constrained lower limits on the starspot temperatures for the crossed starspots.

We included the relative shape of the spot crossings, sliced from each residual Kepler light curve, and scaled their amplitudes as a free parameter in our MCMC analysis with our Spitzer transits. The distributions of relative Spitzer / Kepler spot crossing amplitudes are shown in Figure 1b. The dashed black lines represent the predicted spot crossing amplitude ratio for given spot temperature contrasts. We calculated these temperatures by representing the spots using model, stellar atmospheres at various temperatures[22]. Using $\delta\chi^2$ tests, we indirectly detected spot crossings only at 3.6 micrometres because only these Spitzer observations resulted in positive, bounded photosphere-to-spot temperature contrasts. The 4.5 micrometre Spitzer observations are consistent with zero, or a non-detection at infrared wavelengths. These measurements, especially the non-detections, imply that the starspots crossed during each transit are too hot to mimic water vapour absorption features in the planetary spectrum[12,14]. Our starspot analysis is described in the Methods section along with the distribution of Kepler spot crossing amplitudes for comparison with those observed concurrently with Spitzer.

The activity of HAT-P-11[7,11,12,13] produces variations in the total brightness of the star from spots rotating in and out of view, which will change the band-integrated transit depth measured at different epochs. If the relative stellar brightness at the epoch of each observation is known, then the transit depths can be corrected to a common value. Kepler measured the HAT-P-11's relative brightness during all four Spitzer observations, but not during our WFC3 observation. The unknown stellar brightness during this observation introduced an additional uncertainty in our estimate of the WFC3 transit depth relative to the Spitzer and Kepler observations of ±51ppm. In Figure 2, the offset between the WFC3 spectrum and the best-fit model is, ~98ppm on average.

Figure 2 shows our HAT-P-11b transmission spectrum with Kepler, WFC3, and Spitzer transits combined. We constrain the atmospheric composition using the SCARLET tool, a new evolution of the Bayesian retrieval framework described in previous studies[23,24]. Our primary results are a robust 5.1σ detection of water absorption in the WFC3 data and a 3σ upper limit on the exo-Neptune's atmospheric metallicity of ~700 times solar metallicity (the abundance of heavy elements relative to solar)[5], corresponding to a mean molecular weight of ~10.2 g/mol at the 10 mbar level (Figure 3). Transmission spectra of selected atmospheric models[23,24,25,26] are plotted for a comparison to the observations in Figure 2, with matching symbols in Figure 3. Although the significance of the water vapour detection is unaffected by uncertainties in the stellar activity, because all wavelengths in the water band are measured simultaneously, this uncertainty inhibited placing robust constraints on the methane and carbon dioxide abundance, and therefore the C/O ratio of HAT-P-11b's atmosphere[26].

Figure 3 shows that constraints on the atmospheric metallicity and cloud top pressure are correlated. Atmospheric compositional scenarios along a curved distribution agree with the data at 3σ, spanning a range of atmospheric metallicities from 1 to 700 times solar metallicity. Figure 2 shows that a representative 10,000 times solar (water-dominated) spectrum is robustly excluded by the data. The high mean molecular weight of this atmosphere would not allow the significant water absorption feature observed in the WFC3 band pass.

We found that models with atmospheric metallicities corresponding to solar metallicity require the presence of small particles hazes to simultaneously match the HST and Kepler data points. The fit to the data improves towards higher metallicities, reaching the best-fit value at 190 times solar metallicity. The presence of the water absorption in the WFC3 spectrum required that any cloud deck must be at larger than the 10-mbar pressure level (Figure 3), while the Kepler and Spitzer transit depths similarly impose a lower limit on the cloud top pressure.

The atmospheric and bulk compositions of exoplanets provide important clues to their formation and evolution. Mass and radius alone do not provide unique constraints on the bulk compositions of these planets, which are degenerate for various combinations of rock, ice, and hydrogen gas[27]. By measuring the mean molecular weight of the atmosphere using transmission spectroscopy, we can resolve these degeneracies and provide stronger constraints on the interior compositions of these planets[5,24,27,28]. Observations of water vapour dominate the shape of the infrared spectral features for warm (~1000 K) exoplanets. In contrast, the featureless transmission spectra observed for several, similarly small planets[1,2,3,4,16] ($R_p$ ~ 3–4 $R_\oplus$) imply scattering hazes, clouds, or high mean molecular weights exist in those atmospheres, obscuring absorption features[5,23,24] and limiting our ability to understand their interiors directly[5,24,27]. HAT-P-11b is the smallest and coldest planet with a measured absorption signature through transmission, allowing the estimation of its atmosphere's mean molecular weight, providing new insights into the formation history of this Neptune-mass planet[5,24,27,28,29,30].

**CITATIONS**

1 Knutson, H., *et al.* A Featureless Transmission Spectrum for the Neptune-Mass Exoplanet GJ 436b. *Nature*, **505**, 66-68 (2014)

2 Kreidberg, L., *et al.* Clouds in the Atmosphere of the Super-Earth Exoplanet GJ 1214b. *Nature* **505**, 69-71 (2014)

3 Knutson, H., *et al.* Hubble Space Telescope Near-IR Transmission Spectroscopy of the Super-Earth HD 97658b. http://arxiv.org/abs/1403.4602 (2014).

4 Ehrenreich *et al.* Near-infrared transmission spectrum of the warm-Uranus GJ 3470b with the Wide Field Camera-3 on the Hubble Space Telescope. http://arxiv.org/abs/1405.1056v3 (2014).

5 Moses, J., *et al.* Compositional Diversity in the Atmospheres of Hot Neptunes, with Application to GJ 436b. *Astrophys. J.* **777**, 34 (2013)

6 D'Angelo, G., Durisen, R.H., & Lissauer, J.J. Giant Planet Formation. p. 319 in Exoplanets (ed. S. Seager) in the Space Science Series of the University of Arizona Press (2010)

7 Bakos, G., *et al.* HAT-P-11b: A Super-Neptune Planet Transiting a Bright K Star in the Kepler Field. *Astrophys. J.* **710**, 1724 (2010)

8 Deming, D., *et al.* Infrared Transmission Spectroscopy of the Exoplanets HD 209458b and XO-1b Using the Wide Field Camera-3 on the Hubble Space Telescope. *Astrophys. J.* **774**, 95 (2013)

9 Fazio, G., *et al.* The Infrared Array Camera (IRAC) for the Spitzer Space Telescope. *Astrophys. J.* **154**, 10-17 (2004)

10 Borucki, W., *et al.* Kepler Planet-Detection Mission: Introduction and First Results. *Science* **327**, 977 (2010)

11 Knutson, H., Howard, A., Isaacson, H. A Correlation Between Stellar Activity and Hot Jupiter Emission Spectra. *Astrophys. J.* **720**, 1569 (2010)

12 Deming, D., *et al. Kepler* and Ground-Based Transits of the Exo-Neptune HAT-P-11b. *Astrophys. J.* **740**, 33 (2011)

13 Sanchis-Ojeda, R. & Winn, J. Starspots, Spin-Orbit Misalignment, and Active Latitudes in the HAT-P-11 Exoplanetary System. *Astrophys. J.* **743**, 61(2011)

14 Bernath, P. Water in Sunspots and Stars. *Intern. Astron. U. Symp.* **12**, 70-72 (2002)

15 Berta, Z., *et al.* The GJ1214 Super-Earth System: Stellar Variability, New Transits, and a Search for Additional Planets. *Astrophys. J.* **736**, 35 (2011)

16 Fraine, J.D., *et al.* Spitzer Transits of the Super-Earth GJ1214b and Implications for its Atmosphere. *Astrophys. J.* **765**, 127 (2013)

17 Sing, D., et al. Hubble Space Telescope transmission spectroscopy of the exoplanet HD 189733b: high-altitude atmospheric haze in the optical and near-ultraviolet with STIS. *Mon. Not. R. Astron. Soc.* **416**, 1443 (2011)

18 Patil, A., Huard, D., & Fonnesbeck, C. PyMC: Bayesian Stochastic Modelling in Python. *J. Stat. Soft.* **35**, 4 (2010)

19 Ford, E. Quantifying the Uncertainty in the Orbits of Extrasolar Planets. *Astron. J.* **129**, 1706 (2005)

20 Ford, E. Improving the Efficiency of Markov Chain Monte Carlo for Analyzing the Orbits of Extrasolar Planets. *Astrophys. J.* **642**, 505 (2006)

21 Castelli, F. & Kurucz. R. New Grids of ATLAS9 Model Atmospheres. http://arxiv.org/abs/astro-ph/0405087 (2004)

22 Husser, T.-O., *et al.* A New Extensive Library of PHOENIX Stellar Atmospheres and Synthetic Spectra. *Astron. Astrophys.* **553**, A6 (2013)

23 Benneke, B. & Seager, S. Atmospheric Retrieval for Super-Earths: Uniquely Constraining the Atmospheric Composition with Transmission Spectroscopy. *Astrophys. J.* **753**, 100 (2012)

24 Benneke, B. & Seager, S. How to Distinguish Between Cloudy Mini-Neptunes and Water/Volatile-dominated Super-Earths. *Astrophys. J.* **778**, 153 (2013)

25 Madhusudhan, N., *et al.* A High C/O Ratio and Weak Thermal Inversion in the Atmosphere of Exoplanet WASP-12b. *Nature* **469**, 64 (2011)

26 Madhusudhan, N. C/O Ratio as a Dimension for Characterizing Exoplanetary Atmospheres. *Astrophys. J.* **758**, 36 (2012)

27 Rogers, L. & Seager, S. A Framework for Quantifying the Degeneracies of Exoplanet Interior Compositions. *Astrophys. J.* **712**, 974 (2010)

28 Fortney, J., *et al.* A Framework For Characterizing The Atmospheres Of Low-Mass Low-Density Transiting Planets. *Astrophys. J.* **775**, 80 (2013) 29 Chiang, E. & Laughlin, G. The minimum-mass extrasolar nebula: in situ formation of close-in super-Earths. *Mon. Not. R. Astron. Soc.* 431, 3444-3455 (2013)

30 Hu, R. & Seager S. Photochemistry in Terrestrial Exoplanet Atmospheres. III. Photochemistry and Thermochemistry in Thick Atmospheres on Super Earths and Mini Neptunes. *Astrophys. J.* **784**, 63 (2014)

**Acknowledgements.** J.F., A.J., and N.E. acknowledge support from project IC120009 "Millennium Institute of Astrophysics (MAS)" of the Millennium Science Initiative, Chilean Ministry of Economy, FONDECYT project 1130857 and BASAL CATA PFB-06. N.E. is supported by CONICYT-PCHA/Doctorado Nacional. We thank Peter McCullough for his assistance in the planning and executing of our observations. We are grateful to Ian Crossfield, Laura Kreidberg, and Eric Agol for providing their open source, python code banks on their individual websites. We are also grateful for discussions with Michael Line, Jonathan Fortney, and Julianne Moses about the nature of photochemistry and interior structures. We thank the ATLAS and PHOENIX teams for providing stellar models. We also thank the SciPy and NumPy associations for providing extensive and rigorous numerical routines for an assortment of mathematical and computational techniques.

**Author Contributions.** J.D.F. led the data analysis for this project with contributions from D.D., H.K., N.E., A.J., and A.W. A.W. supplied HST spectral fitting routines and interpretations. N.E. and A.J. supplied python routines for MCMC, wavelet, and transit curve analyses specific to transiting exoplanets. D.D., H.K., N.E., and A.J. provided computational equipment and administration. D.D., N.M., H.K., and K.T. successfully proposed for and provided data from the HST. B.B. and N.M. provided atmospheric models and accompanying fits. B.B. provided atmospheric retrieval analysis with figures and interpretations. N.E. supplied stellar limb-darkening coefficients calculated from both ATLAS and PHOENIX models.

**Author Information.** Reprints and permissions information is available at www.nature.com/reprints. The authors have no competing financial interests to report. Correspondence and requests for materials should be addressed to jfraine@astro.umd.edu or jdfraine@gps.caltech.edu

# FIGURES

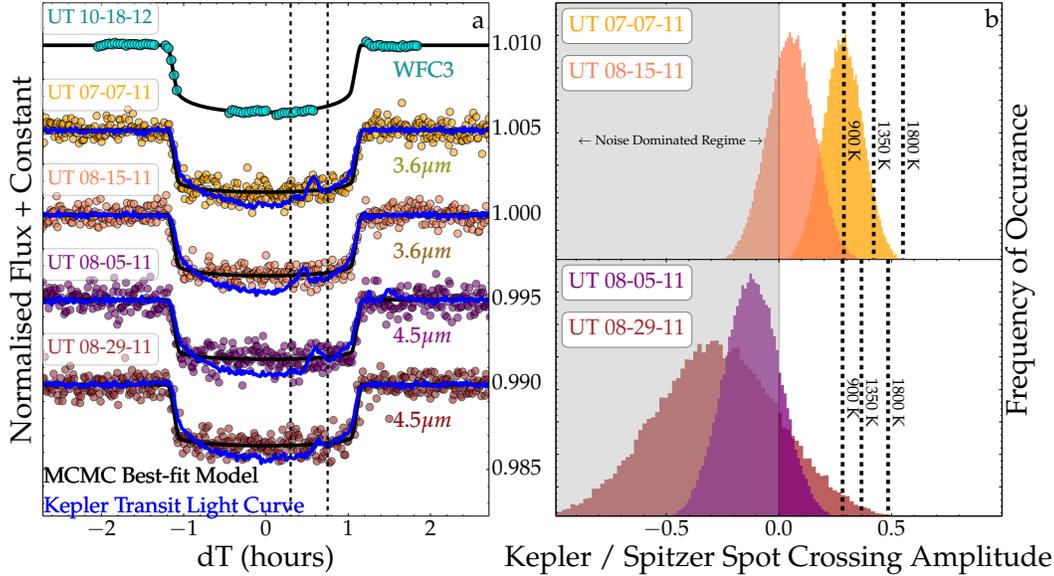

**Figure 1: White-light transit curves and starspot crossing temperature estimates. a,** Transit curves from HST-WFC3 and warm Spitzer, aligned in phase and shifted in flux for clarity. The four warm Spitzer transits at both 3.6 and 4.5 micrometres[9] are binned for illustration. Starspot crossings are seen as deviations near +0.5 hours in the Kepler photometry (dark blue). **b,** We estimated the starspot temperatures by dividing the Spitzer transit residuals by the Kepler transit residuals. The dashed lines represent the photosphere-to-starspot temperatures for three stellar model atmospheres[22]. Water vapour has been detected in sunspots as cool as 3000 K, corresponding to a contrast of ~1800 K here[14]. There is essentially no starspot temperature that can produce sufficiently strong water absorption to mimic our result.

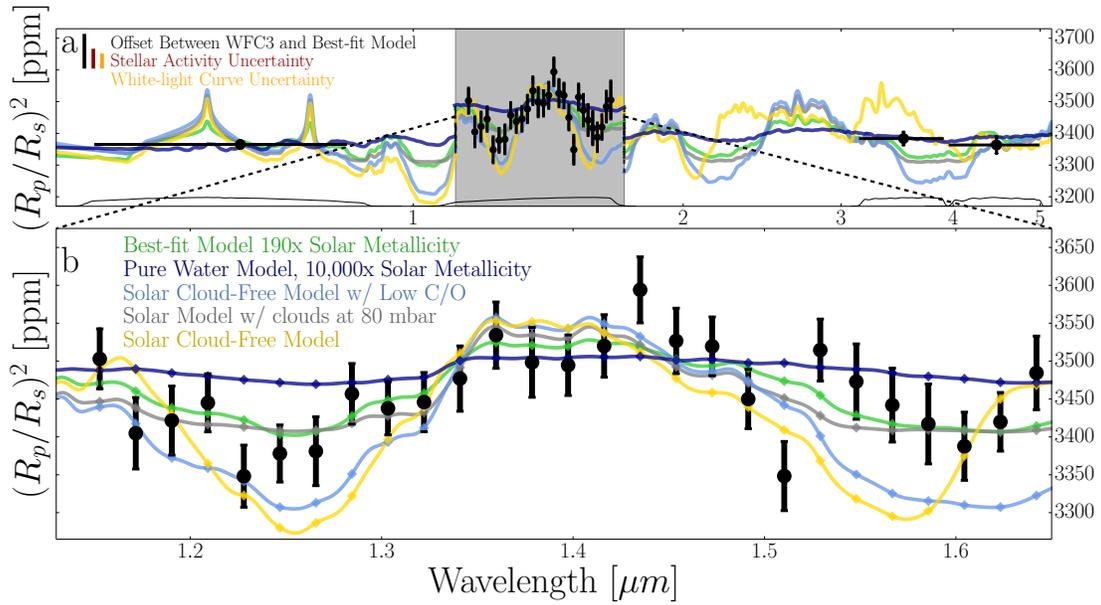

**Figure 2: The transmission spectrum of HAT-P-11b. a**, Our WFC3 observations show a transit depth variations in agreement with a hydrogen-dominated atmosphere. The coloured, solid lines[23,24] correspond to matching markers displayed in Fig. 3. The error bars represent the standard deviations over the uncertainty distributions. High mean molecular mass atmospheres (dark blue line) are ruled out by our observations by >3σ. The WFC3 spectrum was allowed to shift, as a unit, over these uncertainties. **b,** Detailed view of our WFC3 spectrum. For the purposes of visually comparing the spectral significance, we shifted all of the models by 98ppm in the grey region and bottom panel.

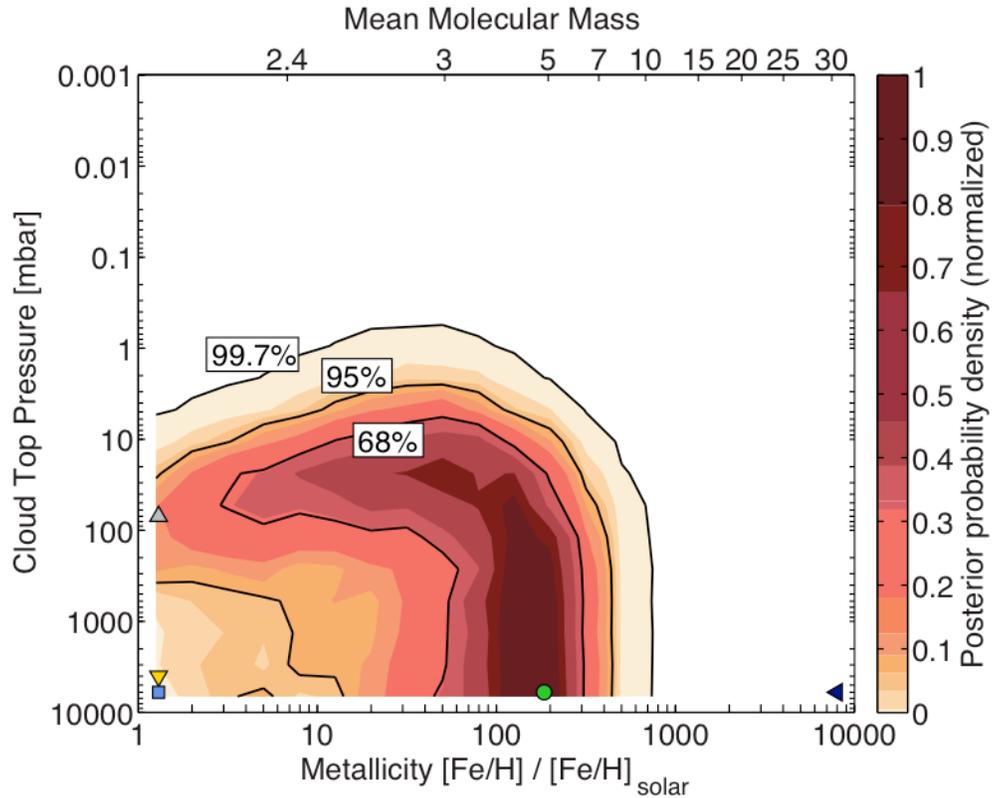

**Figure 3: Spectral retrieval results of our transmission spectrum.** The coloured regions indicate the probability density as a function of metallicity (the abundance of heavy elements relative to solar) and cloud top pressure derived using our Bayesian atmospheric retrieval framework[23,24]. Mean molecular weight was derived for a solar C/O ratio at 10 mbar. Black contours mark the 68%, 95%, and 99.7% Bayesian credible regions. The depth of the observed water feature in the WFC3 spectrum required the presence of a large atmospheric scale height that can self-consistently only be obtained with an atmospheric metallicity below 700 times solar at 3σ (99.7%) confidence. The atmosphere is likely predominately cloud-free at least down to the 1 mbar level. We indicate the matching models plotted in Figure 2 with coloured markers.

| Date | Start Time | End Time | Observatory (Instrument) | Bandpass | Spectral Resolution | Temporal Cadence | Number of Observations |
|---|---|---|---|---|---|---|---|
| **UT Jul 07, 2011** | 23:11:42 | 06:37:52 | warm *Spitzer* (IRAC Channel 1) | 3.6 μm | ~4 | 0.4 s | 62592 |
| **UT Aug 05, 2011** | 07:02:48 | 14:28:58 | warm *Spitzer* (IRAC Channel 2) | 4.5 μm | ~4 | 0.4 s | 58112 |
| **UT Aug 15, 2011** | 01:49:20 | 09:15:30 | warm *Spitzer* (IRAC Channel 1) | 3.6 μm | ~4 | 0.4 s | 52633 |
| **UT Aug 29, 2011** | 17:37:18 | 01:03:28 | warm *Spitzer* (IRAC Channel 2) | 4.5 μm | ~4 | 0.4 s | 62592 |
| **UT Oct 18, 2012** | 17:37:18 | 01:03:28 | Hubble WFC3 (G141) | 1.13-1.64 μm | ~60 – 89 | 123 s | 113 |
| **UT Dec 24, 2012** | 23:56:56 | 03:05:58 | Hubble WFC3 (G141) | 1.13-1.64 μm | ~60 - 89 | 123 s | 99 |

**Table 1: Summary of observations.**
We observed HAT-P-11b during four warm Spitzer observations, two transits at both 3.6 and 4.5 micrometres with the IRAC instrument[9], and two observations using HST WFC3 G141 grism spectrometer, spanning 1.1–1.7 micrometres. Concurrent Kepler observations were retrieved for comparison with our warm Spitzer observations, but were unavailable for our Hubble spectroscopic observations.

# Methods

**Observations**

We observed two transits of HAT-P-11b using Hubble Space Telescope (HST) spectroscopy near 1.4 microns, and we obtained photometry for two transits at 3.6 microns, and two at 4.5 microns, using warm *Spitzer*. We also analysed 208 archival Kepler transits, to assess the effect of starspots and update the optical transit depth[12]; Kepler observed HAT-P-11 during our Spitzer, but not HST, observations.

*Hubble WFC3 Spectroscopy and White Light Photometry*

We observed HAT-P-11 using the HST Wide Field Camera 3 (WFC3) in spatial scan mode (Table 1). We used the G141 grism, with a binned 4-column spectral resolution from 60–89 over the wavelength range 1.1-1.7 microns. Gaps in the HST observations (see Extended Figure 1) occur every ~45 minutes during occultations of the Earth.

We scanned the spectrum in the cross-dispersion direction to maximise efficiency[8,32]. Each scan covered 135 pixels in 44 seconds (~0.3981"/second), yielding ~45,000 electrons per pixel (~70% of saturation). The average photon-limited signal-to-noise ratio (SNR) is ~220 per pixel, integrating to a SNR~2500 per column, for 113 spectral images in transit 1. We were unable to use the second transit because HST's fine guidance sensors lost positional stability, not uncommon in this observing mode[2,8]. This also occurred 8 times during transit 1. We further removed the entire first orbit and the first image of each orbit—a common practice to ameliorate instrumental effects[8,15,33], yielding 72 images for photometric and spectroscopic measurement.

*Warm Spitzer IRAC*

Spitzer transits were critical to establish that starspots on HAT-P-11 are not sufficiently cool to exhibit water absorption masquerading as planetary. Spitzer also provided a long wavelength baseline for the planet's radius, minimising the effect of Rayleigh scattering while remaining sensitive to absorption from carbon-containing molecules such as methane and carbon monoxide.

*Kepler Archival Transits*

We used all 208 archival transits of HAT-P-11 that Kepler observed at ~0.6 microns. The out-of-transit photometry yielded constraints on the disk-integrated activity of the host star (see below). We re-fit the phased & binned Kepler light curve that was analysed previously[12,13], using 4th order limb darkening coefficients to improve the optical radius and geometric parameters of the system. The occurrence of 298 starspot crossings allowed us to characterise the amplitude distribution of spots crossed by HAT-P-11b during transit.

**Data Analysis**

*Limb Darkening Coefficients*

We used a single set of planetary orbital parameters for all analyses, and wavelength-dependent 4-parameter, non-linear limb darkening coefficients[36] (LDCs). For the Kepler, WFC3 band-integrated, and Spitzer bandpasses, we computed the LDCs by integrating stellar model intensities over each instruments' response function. For the 128 individual wavelengths from the WFC3 grism, we used the intersection of the WFC3 response function and a 1-column (4.71 nm) square window centered at each wavelength. We held the computed LDCs constant during subsequent analyses.

We represented HAT-P-11 using an ATLAS model[21] (http://kurucz.harvard.edu/grids.html) having $T_{eff}$ = 4750 K[7]; [M/H]=+0.3; and log(g) = 4.5. To ensure that our exoplanet spectrum is not sensitive to the stellar parameters, we also derived it using LDCs with $T_{eff}$ = 4500 K and $T_{eff}$ = 5000 K. Both the exoplanet spectrum and the white-light (band-integrated) transit depth varied negligibly (~1 and ~5 ppm, respectively) between the three stellar models. Repeating our analysis with both quadratic and three-parameter limb-darkening laws, we found similarly negligible effects (~1 and ~10 ppm).

*Hubble Wide Field Camera 3*

Each WFC3 spatial scan comprised 7 non-destructive reads, with 7.35 seconds of exposure per read. We combined them by subtracting each read from the previous read, applying a spatial mask to the difference, and adding all of the masked differences to an initially blank image to create the spectral frame[1,2,3,8]. We used both edge detection algorithms[37] to determine the edges of each difference image in the scanning direction. We masked all pixels within 20 pixels of this edge in the scanning direction, to keep sky background and other noise from accumulating in the final spectral image. We identified bad pixels in the spectral images using a median filter with a 4-σ threshold, over a 7-pixel window, and assigned bad pixel values to the median of the window. Extended Figure 2 shows a spectral frame from the first visit, displaying the spectral (dispersion) and spatial (cross-dispersion) dimensions. The curves in Extended Figure 2b show the averaged, column-integrated spectral template (red), before (top) and after (bottom) being fit to the example spectrum (blue) for both the wavelength solution and the white-light photometry as a function of time.

We calculated the wavelength flat field calibration using standard procedures[31]. We fit a 2D Gaussian to the spectral images, and found that the spectra shifted by at most 1.12 pixels in the wavelength direction; we corrected this shift during the template fitting described below. The final, column integrated spectra were derived by dividing each raw spectral image by its corresponding flat-field and subtracting the per-column background values, then summing the detector in the cross-dispersion direction (down the columns).

The sky background was calculated per-column as the median of the portion of the spectral image not scanned by the instrument. In Extended Figure 2a, this corresponds to the blue regions above and below the red/orange spectral information. The background values varied < 3% from 1.1 to 1.7 microns, but were uncorrelated[37] with the resultant planetary spectrum.

To derive the WFC3 spectrum for HAT-P-11b, we used the established technique[1,3,8,33] of spectral template fitting. We formed the spectral template by averaging the out-of-transit spectra, and fit it to individual grism spectra in both wavelength and amplitude[37], using both Levenberg-Marquardt and spline interpolation algorithms[37]. The fitted amplitudes as a function of time yield the band-integrated white light curve (WLC) (see Extended Figure 1). The WLC defines the average transit depth over the total WFC3 bandpass.

WFC3 Exponential and Linear Baselines:

The raw WLC contained both the transit and instrument effects in the form of exponential ramps[1,2,8,15] over each orbit, and a linear trend over each visit. We simultaneously fit for exponential parameters as a function of HST's orbital phase, $E_R(\theta;A,S)$, a linear trend as a function of time, $L(t;m,b)$, an additive offset for the second half of the in-transit data $O(t_i;\theta,O_o)$ (see below), with an analytic transit light curve[38], $MA(t; p, T_c, P, b, a, u_1, u_2)$:

$E_R(\theta; A, S) = 1 + A * \exp\{ S * (\theta_i - \theta_{min})\}$ (2a)
$O(t_i, \theta; O_o ) = O_o * \delta(\theta_i - \theta_2) * \chi(t_i)$ (2b)
$L(t; m,b) = m*(t_i - t_{min}) + b$ (2c)
$Model = MA(t; p, T_c, P, b, a, u_1, u_2)* E_R(\theta; A_o, S_o)* L(t; m,b) + O(t_i; \theta, O_o )$ (2d)

where $\theta$ represents HST's orbital phase, $\{A, S\}$ are the exponential amplitude and scale factor, and $\{m, b\}$ are the slope and intercept of the linear function. $O(t_i; \theta, O_o)$ is a small (~100 ± 50 ppm) step function, $\chi(t_i)$, correcting an unexplained offset in the band-integrated photometry (see Extended Figure 1 between phase ~0.2–0.7). The offset is likely related to a small shift in the position of the spectrum that does not occur in other HST orbits. Including the offset in our model improved the WLC fits significantly, without degrading the Bayesian Information Criterion[39,40] (BIC), or altering the significance of the water detection.

We tested four different models for WFC3's WLC exponential baseline[8,15,39], and selected among them based on the BIC. The BIC for our adopted model differed only slightly from optimum ($\Delta BIC << 2$)[39,40], but the transit depth was more physically realistic compared to atmospheric models. WFC3 WLCs are known to have noticeable red noise[8,33,1], so we also implemented a wavelet analysis to include both the white noise ($\sigma_w$ = 12.81ppm) and the red noise ($\sigma_r$ = 61.89 ppm) components of the residuals into our final uncertainties[41].

We use a Markov Chain Monte Carlo (MCMC) procedure[18] to simultaneously fit the transit and instrument parameters, thereby incorporating correlations between parameters into our reported uncertainties. Extended Figure 3 compares the posteriors graphically. The Pearson correlation coefficient over the each parameter found the correlations to be insignificant ($\max_{i,j}(P) < \pm 0.10 \; \forall \; i,j \in$ {fitted parameters}).

HST WFC3 Exoplanet Spectrum Derivation:

We calculated the planetary spectrum differentially relative to the WLC, by dividing the spectral template, $S_T(\lambda)$, into each individual spectrum $S(\lambda, t_i)$, allowing for small wavelength shifts (< 1.12 pixels). The planetary spectrum, $P(\lambda)$, is derived from the normalised residuals, or *differential light curves*:

$$DLC(\lambda, t_i) = (\; S(\lambda, t_i) - S_T(\lambda, t_i)\;) / S_T(\lambda, t_i). \tag{3}$$

We fit the $DLC(\lambda, t_i)$, with *differential analytic light curves*, $DALC(\lambda, t_i)$, by renormalising the analytic WLCs[38], $MA(t_i)$:

$$DALC(\lambda, t_i) = (\; MA(t_i) - \max_t\{MA(t_i)\}\;)/(\; \max_t\{MA(t_i)\} - \min_t\{MA(t_i)\}\;) + \max_t\{MA(t_i)\} \tag{4}$$

We fit the normalisation amplitude, $P(\lambda)$, of the DALC's, simultaneously with wavelength-dependent linear trends, using linear matrix inversion

$$DLC(\lambda, t_i) = P(\lambda) * DALC(\lambda, t_i) + m_\lambda * (t_i - t_{min}) + b_\lambda \tag{5}$$

where the minima and maxima are taken over the time domain. Extended Figure 4 shows all 32 DLCs (blue-to-red) and DALCs (black) ranging from 1.167 to 1.675 microns with 18nm spacing in wavelength. We used the linear matrix fits as initial conditions for MCMCs to probe posterior distributions per wavelength, forming the final planetary spectrum $P(\lambda)$ shown in Figure 2b. The DALCs were modified to include wavelength-dependent limb darkening described above. Our analysis also included smoothing in wavelength with a triangle function to reduce the effect of known, spectral undersampling with WFC3 data[1,3,8]. The full width half maximum of the smoothing triangle was 4 columns, resulting in 32 DLCs at a spectral resolution of R~75.

*Warm Spitzer IRAC*

We performed aperture photometry on the Spitzer images (Table 1), after subtracting a background value determined as a median of the pixel values well away from the stellar image. We applied a newly developed base vector analysis to the subsequent photometry in order to decorrelate the well-known Spitzer intrapixel effect[16,35,42,43,44]. We tested this decorrelation method on both published and unpublished Spitzer exoplanet eclipses, and found that it consistently reduces both the BIC and red-noise compared to previous methods[35,42,43,44]. Our base vector algorithm fits linear coefficients to the 9 pixel values centered at and surrounding the stellar PSF[16] over time:

$$\text{Flux}(t) = a^{-1}(t) \sum_n a_n * \text{pixel}_n(t); n = \{1,...,9\}, \tag{6}$$

where $a(t)$ is the normalisation for each 3x3-pixel box as a function of time. Our Spitzer fitting simultaneously solved for the transit depth and amplitudes of starspot crossings, by scaling the more precise (and strictly simultaneous) Kepler spot crossings as described below.

### Kepler Archival Transits

We used the Kepler data for two completely independent purposes. The first purpose was to derive an improved optical transit depth and geometric parameters. The second purpose was to characterize the nature of starspots on HAT-P-11.

### Improving the Optical Transit Depth

To improve the optical transit depth, we improved on the previous analysis[12], using all 4 years of Kepler data Q0-Q16, incorporating 4th-order limb darkening in the analysis. We detrended the stellar variations by fitting a straight line to the data within one transit duration of 1st and 4th contacts. After dividing by these straight lines, we phase folded all 208 short cadence Kepler archival transits into a single, very high precision transit having 60 second time resolution. We fit an analytic transit model[38] to this phase folded light curve, including the LDCs described above, and geometric parameters from radial-velocity measurements[45]. This determined both the transit depth and the geometric parameters (e.g., impact parameter) to high precision. All of the values were within 1-σ of the original analysis[12], but now followed the same limb darkening model as all of our data sets.

### The Nature of Starspots on HAT-P-11

Our second analysis of the Kepler data characterized the nature of starspots that occur on the disk of HAT-P-11. We use the properties of spots crossed by the planet during transit to verify that un-crossed spots cannot mimic water absorption in the planetary spectrum. This involves determining the relative temperature (spot vs. photosphere) of the spots crossed during the Spitzer transits, and also demonstrating that those spots are representative of the uncrossed spots.

To determine the relative amplitude (Spitzer vs. Kepler) of the spot crossings, we first measured their profiles in the Kepler transits as the residual of the individual

Kepler light curves relative to the phased & binned model. We include the Kepler spot crossing profile as a term in the MCMC fitting[18,19,20] to the Spitzer transits, fitting the spot crossing amplitude simultaneously with the transit parameters. Negative ratios are statistically possible here because of the low SNR of the spot crossing profiles. We allowed our MCMC chains to probe these unphysically negative values of spot crossing amplitudes. In Figure 2b, the grey region represents the noise-dominated regime and the white region represents the physically relevant regime. Because the spot area crossed by the planet is the same for the simultaneous Spitzer and Kepler transits, the relative amplitudes can be converted to relative temperatures in Figure 1b.

To establish that the crossed spots are typical of the uncrossed ones, we fit a Gaussian profile to each of the 298 Kepler starspot crossing profiles-- during the 208 Kepler transits observed over 4 years-- to determine the full distribution of the spot crossing amplitudes (see Extended Figure 5), discussed below.

*Constraining the Significance of Water Vapour in Starspots*

The amplitude of the starspot effect that mimics exoplanetary absorption is given as $f\varepsilon\delta$[8], where $f$ is the fractional coverage of spots on the stellar disk, $\varepsilon$ is the depth of the transit, and $\delta$ is the water vapor absorption line depths (relative to the spectral continuum) in the spatially-resolved starspot spectrum, at the observed spectral resolution. Because the planet's orbit is not synchronous with the star's rotation, spot crossings in the Kepler data were used to estimate a flux deficit of $0.0179$[12]. Accounting for the (small) optical intensity of dark starspots lead to $f=0.02$. $\varepsilon=0.0036$ is measured directly by our WFC3 photometry (Figure 1a). Since we cannot obtain spatially resolved spectra of the starspots on HAT-P-11, we here estimate $\delta$ from Phoenix model stellar atmospheres[22], using a relative temperature constraint for the starspots based on Figure 1b.

To establish that the spots occulted in our simultaneous Kepler and Spitzer photometry are typical of HAT-P-11, we plot the distribution of spot crossing amplitudes (Extended Figure 5) over the entire set of Kepler photometry for HAT-P-11 (Extended Figure 6).

Spots that cover the disk of the star cause a brightness variation as the star rotates. This variation has an approximately 2% peak-to-peak modulation, consistent with the spot coverage inferred from our previous Kepler study[12]. The times of our Spitzer and WFC3 observations are indicated on Extended Figure 6 by blue and red lines, respectively. Similarly, the spot crossing amplitudes during our Spitzer observations are identified on Extended Figure 5 as the colored, dashed lines. From this we concluded that the total effect of spots on the disk of HAT-P-11 is approximately the same during our observations as during other times, and that spots crossed during our Spitzer observations are typical of the unocculted spots during our WFC3 observations.

Next we must determine δ, the depth of the 1.4 micron water absorption feature in the spectrum of starspots. We approximate the starspot spectrum as equivalent to a star of the same abundance and surface gravity as HAT-P-11, but having a cooler temperature. We examined Phoenix model atmospheres that are enriched in oxygen by +0.3 in the log of abundance[22], and we convolved their spectra to the resolution of WFC3. Even in the extreme case with a temperature contrast of 1800 K (Figure 1b), the Phoenix spectrum shows that δ ~ 0.024 at 1.4 microns, yielding $f\varepsilon\delta$ < 2 ppm, which is two orders of magnitudes less than the absorption we derived for the exoplanetary atmosphere. Because $f$ and ε are small, there is essentially no starspot temperature that can produce sufficiently strong water absorption to mimic our result, given our inferred values for $f$ and ε.

Self-Consistent Atmospheric Retrieval for Exoplanets

We interpret the observed transmission spectrum using a new variant of our atmospheric retrieval framework described in previous studies[23,24]. The new SCARLET framework combines a self-consistent, line-by-line atmospheric forward model with the nested sampling technique to efficiently compute the joint posterior probability distribution of the atmospheric parameters. We probe the multidimensional parameter space spanned by the metallicity (i.e. the overall abundance of heavy elements), the carbon-to-oxygen ratio (C/O), the cloud top pressure, the planetary radius at the 1 bar level, and the planetary Bond albedo.

For a given set of parameters, the atmospheric forward model self-consistently computes the molecular abundances in chemical equilibrium and the temperature pressure profile in radiative-convective equilibrium. Line-by-line radiative transfer based on pre-calculated opacity look-up tables enables us to accurately model molecular absorption for the entire range of compositions. Rayleigh scattering is included using the two-stream approximation. In this study, we included clouds as a grey opacity source that cuts of the transmission of starlight below the parameterized cloud top pressure.

We included the planetary Bond albedo as a free parameter to capture the uncertainty in the atmospheric composition introduced by the unknown albedo. For a given atmospheric composition, the Bond albedo introduced the dominant uncertainty in the planetary temperature profile, which (via the scale height) affects the relation between observed depths of the absorption features and the mean molecular mass.

The nested sampling algorithm repeatedly invoked the atmospheric forward model to probe the agreement between model spectra and the observational data throughout the multidimensional parameter space. In total, several $10^4$ self-consistent, line-by-line atmospheric models are computed. The algorithm is initiated by randomly sampling 1000 active samples within the full multidimensional parameter space. The active samples then iteratively migrate towards the regions of high likelihood by replacing the active sample with the lowest likelihood, i.e. the worst fit to the observations, by a new better fitting random sample[24]. Convergence is obtained once the logarithm of the Bayesian evidence, Z, computed from the active sample no longer changes by more than $\Delta(\log(Z))=0.0001$. The algorithm is robust to multimodal posterior distributions and highly elongated curving degeneracies frequently encountered in exoplanet atmospheric retrieval studies[24].

## CITATIONS

31 Rajan, A., *et al.* WFC3 Data Handbook 2011. (Space Telescope Science Institute, 2011)

32 McCullough, P. M. & MacKenty, J. Considerations for using Spatial Scans with WFC3. Instrument Science Report WFC3 2012–8 (Space Telescope Science Institute, 2012).

33 Wilkins, A., *et al*. The Emergent 1.1-1.7 μm Spectrum of the Exoplanet Corot-2b as Measured Using The Hubble Space Telescope. *Astrophys. J.* **783,** (2014)

34 Todorov, K., *et al.* Warm Spitzer Observations of Three Hot Exoplanets: XO-4b, HAT-P-6b, and HAT-P-8b. *Astrophys. J.* **746,** (2012)

35 Lewis, N., *et al.* Orbital Phase Variations of the Eccentric Giant Planet HAT-P-2b. *Astrophys. J.* **766,** (2013)

36 Claret, A. A new non-linear limb-darkening law for LTE stellar atmosphere models. Calculations for -5.0 <= log[M/H] <= +1, 2000 K <= $T_{eff}$ <= 50000 K at several surface gravities. *A&A* **363,** (2000)

37 Oliphant, T. Python for Scientific Computing. *CS&E* **9** (2007)

38 Mandel, K. & Agol, E. Analytic Light Curves for Planetary Transit Searches. *Astrophys. J.* **580,** (2002)

39 Mandell, A., *et al.* Exoplanet Transit Spectroscopy Using WFC3: Wasp-12 b, Wasp-17 b, And Wasp-19 b. *Astrophys. J.* **799**, 128 (2013)

40 Kass, R. & Raftery, A. Bayes Factors. *J. Am. Stat. Assoc.* **90,** 430 (1995)

41 Carter, J. A., & Winn, J. N. Parameter Estimation from Time-series Data with Correlated Errors: A Wavelet-based Method and its Application to Transit Light Curves. *Astrophys. J.* **704,** 51 (2009) 42 Knutson, A. *et al.* The 3.6-8.0 μm Broadband Emission Spectrum of HD 209458b: Evidence for an Atmospheric Temperature Inversion. *Astrophys. J.* **673,** (2008)

43 Ballard, S. *et al.* A Search for a Sub-Earth-Sized Companion to GJ 436 and a Novel Method to Calibrate Warm Spitzer IRAC Observations *PASP* **122,** 2010

44 Knutson, H. *et al.* 3.6 and 4.5 μm Phase Curves and Evidence for Non-equilibrium Chemistry in the Atmosphere of Extrasolar Planet HD 189733b. *Astrophys. J.* **754, (**2012)

45 Knutson, H., *et al.* Friends of Hot Jupiters. I. A Radial Velocity Search for Massive, Long-period Companions to Close-in Gas Giant Planets. *Astrophys. J.*, **785**, 126 (2014)

**EXTENDED FIGURES**

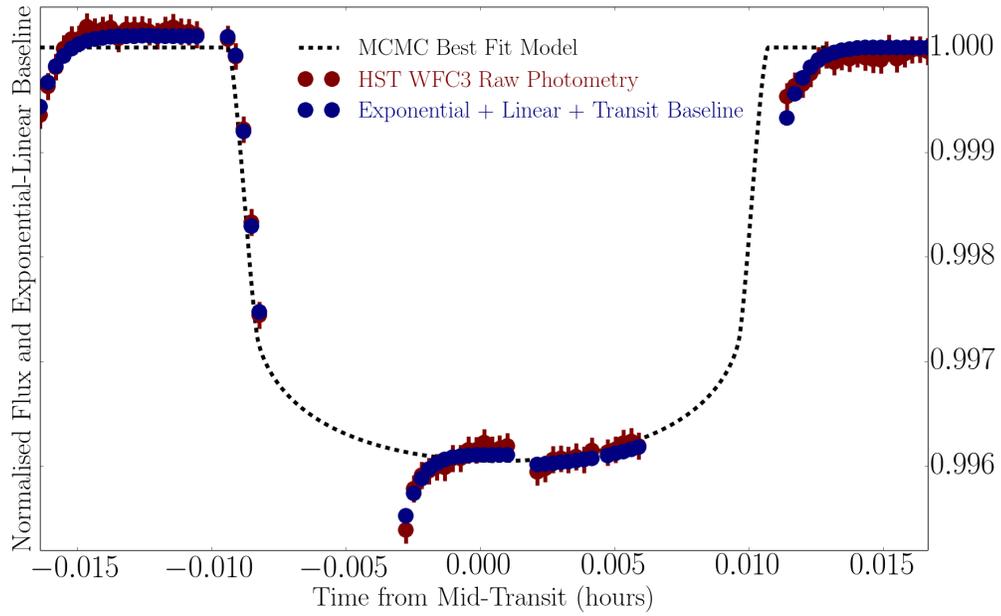

**Extended Figure 1: HST White Light Curve with exponential ramp effects.** The gaps resulted when HAT-P-11 was occulted by the Earth during Hubble's ~96 minute orbit. We decorrelated the ramp effect by fitting an average, 2-parameter (scale and amplitude) exponential function over time.

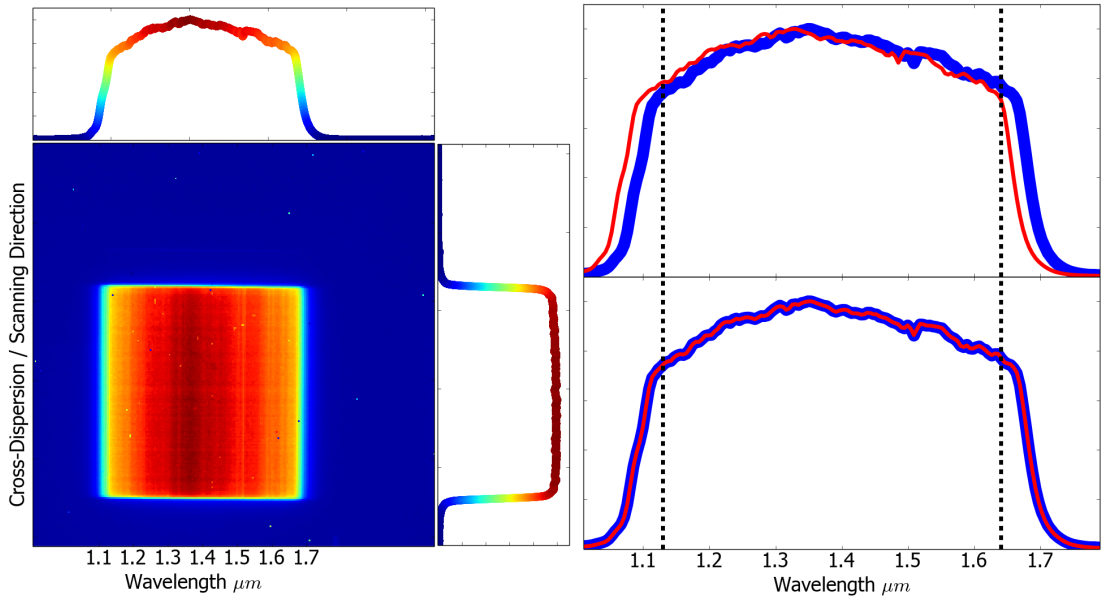

**Extended Figure 2: An example of WFC3 scanning mode observation spectral images. a,** Example spatial scan spectral image with the normalised summations in the dispersion (upper) and cross-dispersion or scanning (right) directions. **b,** Integrated spectrum (blue) and spectral template (red) before (top) and after (bottom) fitting; the amplitudes and colors are normalised to 1.0.

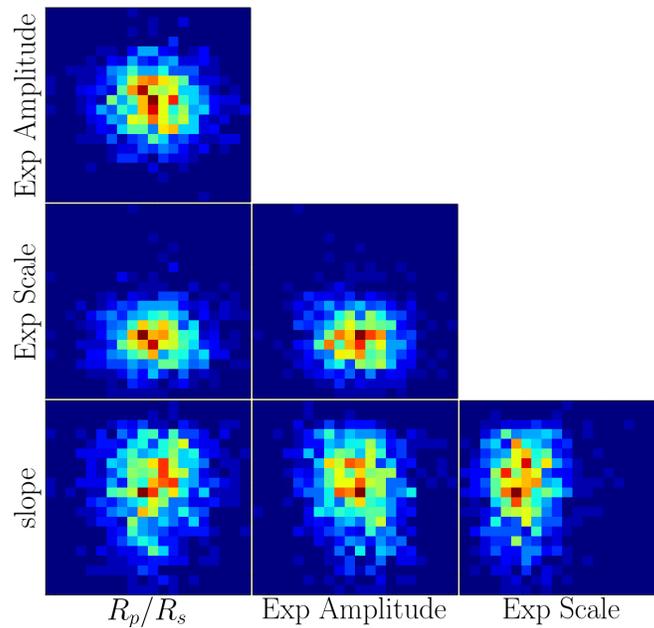

**Extended Figure 3: Correlations between all fitted parameters for our HST WFC3 White Light Curve.** We calculated the Pearson correlation coefficient over the posteriors of each parameter, and found the correlations to be small (< ±0.10), or in most cases negligible (< ±0.01). Blue represents regions of lesser posterior density and the Red represents regions of greater posterior density, with green and yellow in the middle.

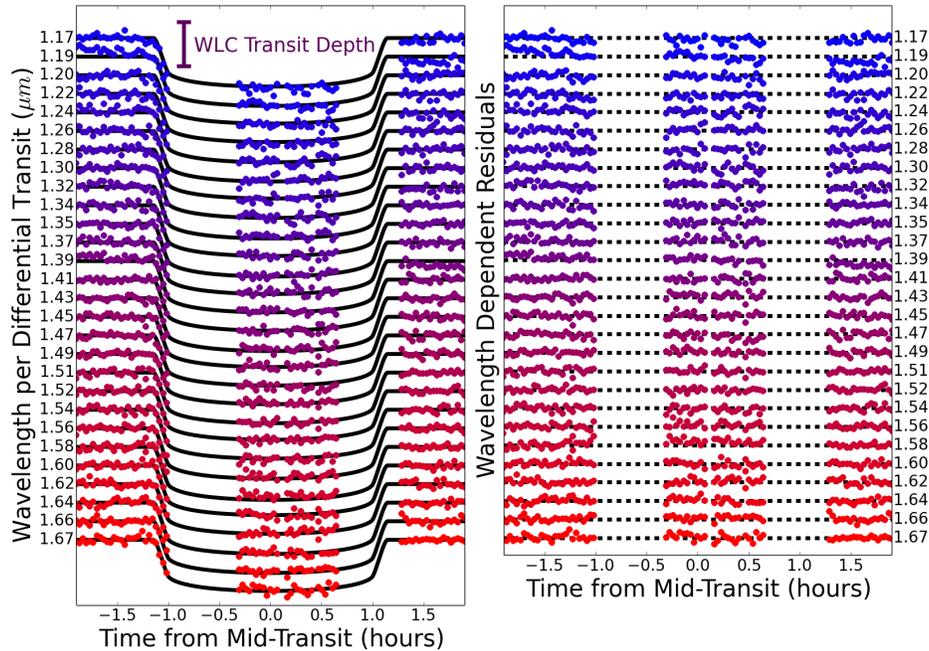

**Extended Figure 4: The full collection of 32 Wavelength Dependent Light Curves (Blue/Red) and Wavelength Dependent Analytic Light Curves.** The black lines represent the best-fit transit light curves over the wavelength range from 1.1 to 1.7 microns, with 18nm spacing in wavelength. The colour is associated with the wavelength, ranging from blue (1.17 microns) to red (1.67 microns). The light curves have been shifted for display purposes only. The differential light curves were fit with differential analytic transit curves to derive the planetary spectrum seen in Figure 1.

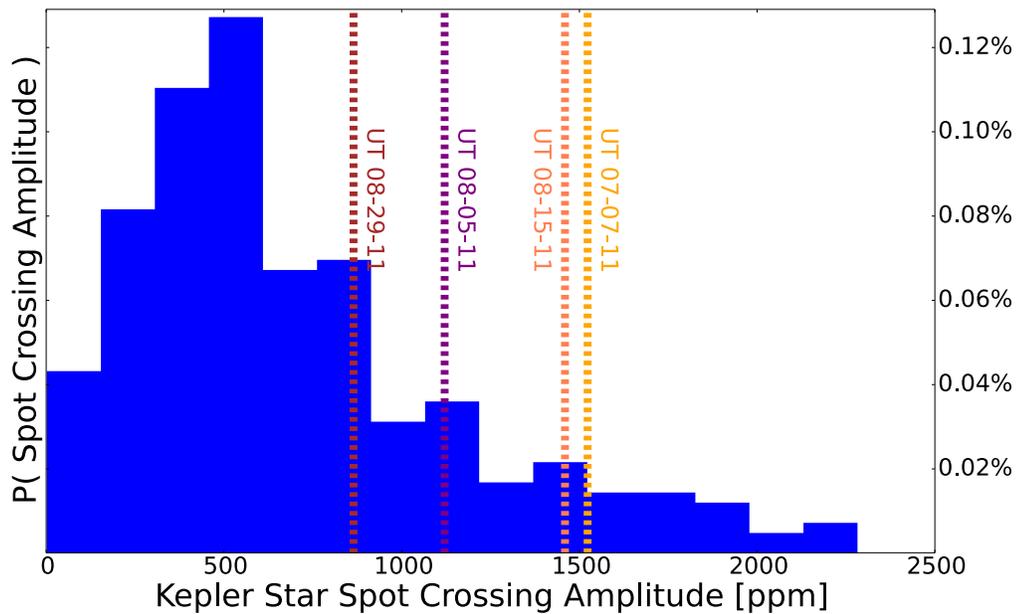

**Extended Figure 5: The distribution of Kepler starspot crossing anomalies.** We fit a Gaussian profile to each of the 298 spot crossings observed during the 208 transits observed by Kepler. Here we show the distribution of starspot amplitudes, calculated as the height - baseline of the fitted Gaussian profile in parts-per-million. The dashed lines represent the starspot crossing amplitudes observed during our 4 concurrent Spitzer observations. In particular, note that all 4 spot crossings with concurrent Spitzer observations are on the larger end of the distribution. In addition, the spot crossing on UT 07-07-11, the largest spot crossing feature observed during our concurrent Spitzer observations, crossed a spot with ΔT ~ 900K.

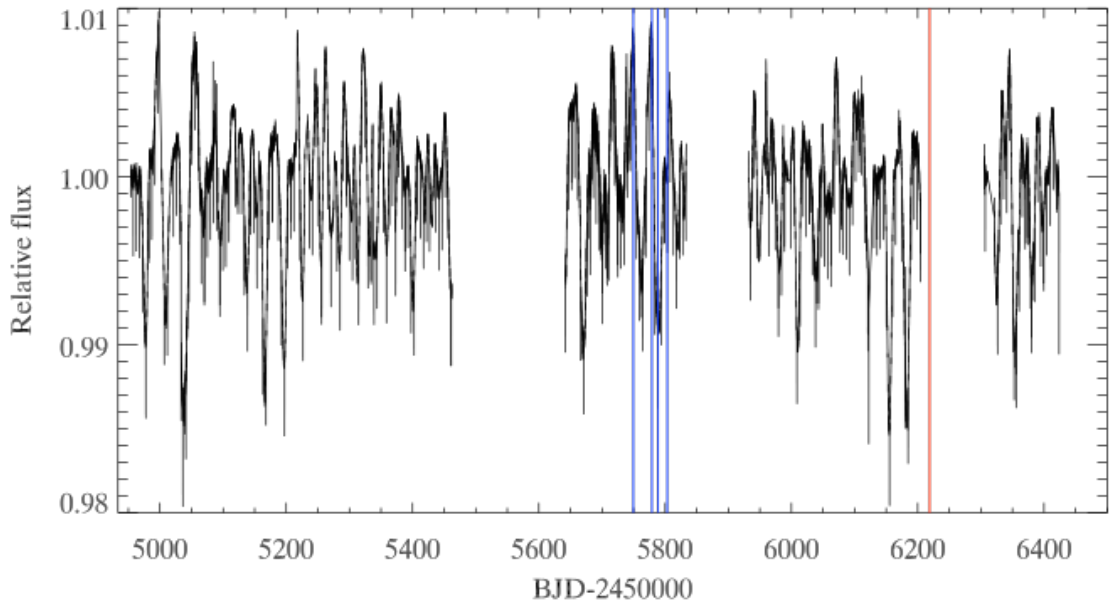

**Extended Figure 6: The full Kepler light curve for Q0 - Q16 (~4 years of short cadence).** The modulation has an approximately 2% peak-to-peak modulation, consistent with the spot coverage inferred from our previous Kepler study. The times of our Spitzer observations are marked with vertical blue lines. The time of our HST WFC3 observation, included in this analysis, are marked with vertical red lines.

| Wavelength [microns] | Transit Depths [ppm] | Transit Depth Uncertainties [ppm] | Bandpass- / Bin-width [microns] |
|---|---|---|---|
| 0.642 ♭ | 3365 | 8 | 0.2415 |
| 1.153 | 3502 | 40 | 0.0094 |
| 1.172 | 3407 | 47 | 0.0094 |
| 1.190 | 3421 | 46 | 0.0094 |
| 1.209 | 3445 | 38 | 0.0094 |
| 1.228 | 3350 | 41 | 0.0094 |
| 1.247 | 3377 | 38 | 0.0094 |
| 1.266 | 3380 | 45 | 0.0094 |
| 1.284 | 3457 | 40 | 0.0094 |
| 1.303 | 3436 | 35 | 0.0094 |
| 1.322 | 3448 | 39 | 0.0094 |
| 1.341 | 3476 | 43 | 0.0094 |
| 1.360 | 3536 | 44 | 0.0094 |
| 1.379 | 3499 | 46 | 0.0094 |
| 1.397 | 3498 | 40 | 0.0094 |
| 1.416 | 3524 | 41 | 0.0094 |
| 1.435 | 3591 | 44 | 0.0094 |
| 1.454 | 3524 | 43 | 0.0094 |
| 1.473 | 3520 | 39 | 0.0094 |
| 1.492 | 3447 | 39 | 0.0094 |
| 1.510 | 3344 | 45 | 0.0094 |
| 1.529 | 3513 | 41 | 0.0094 |
| 1.548 | 3471 | 50 | 0.0094 |
| 1.567 | 3438 | 49 | 0.0094 |
| 1.586 | 3414 | 53 | 0.0094 |
| 1.604 | 3383 | 45 | 0.0094 |
| 1.623 | 3415 | 38 | 0.0094 |
| 1.642 | 3480 | 48 | 0.0094 |
| 1.661 | 3498 | 60 | 0.0094 |
| 1.680 | 3376 | 74 | 0.0094 |
| 3.521* | 3384 | 20 | 0.3685 |
| 4.471* | 3363 | 27 | 0.5021 |
| 3.521 | 3421 | 29 | 0.3685 |
| 3.521 | 3347 | 29 | 0.3685 |
| 4.471 | 3321 | 38 | 0.5021 |
| 4.471 | 3407 | 37 | 0.5021 |

**Extended Table 1: Transit depths as a function of wavelength for Kepler, HST WFC3, Spitzer IRAC1, and Spitzer IRAC2.**
♭ Kepler transit depth determined from all 208 phased and binned Kepler transits
* Weighted mean of the two other independent Spitzer transits for each channel

|  | Kepler | HST WFC3 | Spitzer IRAC1 UT 07-07-11 | Spitzer IRAC1 UT 08-15-11 | Spitzer IRAC2 UT 08-05-11 | Spitzer IRAC2 UT 08-29-11 |
|---|---|---|---|---|---|---|
| Wavelength (μm) | 0.641 | 1.419 | 3.521 | 3.521 | 4.471 | 4.471 |
| Period (days)[a] | 4.8878018 ± 7.1 x 10$^{-6}$ | -- | -- | -- | -- | -- |
| $T_{center}$ (J.Days) | 54811.1786727 ± 0.0000082 | 54811.1785254 ± 0.0000663 | 54811.1791713 ± 0.0001272 | 54811.1788248 ± 0.0001239 | 54811.1790119 ± 0.0001541 | 54811.1789256 ± 0.0001591 |
| Inclination (°)[b] | 89.549 ± 0.114 | -- | -- | -- | -- | -- |
| b (impact parameter)[b] | 0.135 ± 0.034 | -- | -- | -- | -- | -- |
| $a / R_s$[b] | 17.125 ± 0.060 | -- | -- | -- | -- | -- |
| $R_p / R_s$ | 0.05852 ± 0.00007 | 0.05887 ± 0.00025 | 0.05849 ± 0.00025 | 0.05785 ± 0.00033 | 0.05762 ± 0.00033 | 0.05837 ± 0.00032 |
| Transit Depth[c] (ppm) | 3424 ± 8 | 3466 ± 29 | 3421 ± 29 | 3347 ± 29 | 3321 ± 38 | 3407 ± 37 |
| $c_1$[d] | 0.7547 ± -- | 0.6718 ± -- | 0.5750 ± -- | 0.5750 ± -- | 0.6094 ± -- | 0.6094 ± -- |
| $c_2$[d] | -0.9164 ± -- | -0.1618 ± -- | -0.3825 ± -- | -0.3825 ± -- | -0.7325 ± -- | -0.7325 ± -- |
| $c_3$[d] | 1.6411 ± -- | 0.2855 ± -- | 0.3112 ± -- | 0.3112 ± -- | 0.7237 ± -- | 0.7237 ± -- |
| $c_4$[d] | -0.6328 ± -- | -0.1551 ± -- | -0.1099 ± -- | -0.1099 ± -- | -0.2666 ± -- | -0.2666 ± -- |
| Eccentricity[e] | 0.232 ± 0.054 | -- | -- | -- | -- | -- |
| Longitude of Periastron[e] | 7° ± 25° | -- | -- | -- | -- | -- |

**Extended Table 2: The system and planetary parameters of HAT-P-11b.**
-- Value held constant for all MCMC chains
[a]Deming et al. 2011
[b]Derived from MCMC posteriors over our phased & binned Kepler transit.
[c]Transit depths are uncorrected for stellar activity.
[d]Derived from ATLAS models
[e]Knutson et al 2014c, Astrophys. J., submitted